\documentclass[%
reprint,
 amsmath,amssymb,
 aps,
prb,
]{revtex4-2}
\usepackage{graphicx, caption}
\usepackage{amsmath,bm}
\usepackage{graphicx}
\usepackage{chemformula}
\usepackage{dcolumn}
\usepackage{bm}
\usepackage[colorlinks=true,citecolor=blue,linkcolor=blue,urlcolor  = blue, breaklinks=true]{hyperref}

\begin{document}

\preprint{APS/PRB}

\title{Highly tunable spin Hall magnetoresistance in room-temperature magnetoelectric multiferroic, \ch{Sr_3Co_2Fe_{24}O_{41}}$\vert$\ch{Pt} hybrids}

\author{Aditya A. Wagh}
\email[]{adityawagh@iisc.ac.in}
\affiliation{%
 Department of Physics, Indian Institute of Science, Bangalore, INDIA 
}%

\author{Priyanka Garg}
\email[]{priyankagarg@iisc.ac.in}
\affiliation{%
 Department of Physics, Indian Institute of Science, Bangalore, INDIA 
}%

\author{Kingshuk Mallik}
\affiliation{%
 Department of Physics, Indian Institute of Science, Bangalore, INDIA 
}%

\author{Suja Elizabeth}
\affiliation{%
 Department of Physics, Indian Institute of Science, Bangalore, INDIA 
 }

\author{P. S. Anil Kumar}
 \email[]{anil@iisc.ac.in}
\affiliation{%
 Department of Physics, Indian Institute of Science, Bangalore, INDIA 
}%

\begin{abstract}

We present spin transport studies on a low-field, room-temperature magnetoelectric multiferroic polycrystalline \ch{Sr_3Co_2Fe_{24}O_{41}}$\vert$\ch{Pt} heterostructure wherein a highly tunable transverse conical magnetic phase is responsible for static and dynamic magnetoelectric coupling. We measured angular dependence of spin Hall magnetoresistance (SMR) at constant magnetic fields ($H$) in the range  of 50 to 100 kOe. Application of field below the critical value ($H_{\text{c1}} \approx 2.5$ kOe), yielded negative SMR and the $H$-evolution of normalized SMR ($(\Delta R / R_{0}) \times 100 \%$) exhibited a negative gradient. Further, an increase in the $H$ resulted in the positive slope of $(\Delta R / R_{0}) \times 100 \%$ Vs. $H$ and later at higher $H$ around 14 kOe, a crossover from negative to positive SMR was observed. We employed a simple model for estimating the equilibrium magnetic configuration and computed the SMR modulation at various values of $H$. We argue that the tilting of the cone is dominant and in turn responsible for the observed nature of SMR below 2.5 kOe while, the closing of the cone-angle is pronounced at higher fields causing a reversal in sign of the SMR from negative to positive. Importantly, SMR experiments revealed that a change in the helicity with a reversal of the magnetic field has no influence on the observed SMR. Longitudinal spin Seebeck effect (LSSE) signal was measured to be $\approx 500$ nV at 280 K, under application of thermal gradient, $\Delta T = 23$ K and field, 60 kOe. The observed LSSE signal, originating from pure magnon spin current, showed a similar $H$-dependent behavior as that of the magnetization of \ch{Sr_3Co_2Fe_{24}O_{41}}. Our detailed spin transport studies on polycrystalline \ch{Sr_3Co_2Fe_{24}O_{41}}$\vert$\ch{Pt} heterostructure demonstrate high tunability of the amplitude and the sign of the SMR, highlighting its potential for novel spintronic devices such as SMR-based spin valves and voltage-controlled spin transport devices.

\end{abstract}

\maketitle

\section{\label{sec:level1} INTRODUCTION}

Spin transport across magnetic insulator$\vert$heavy metal (MI$\vert$HM) interfaces have aroused immense interest in the spintronics community in recent years \cite{nakayama2013spin, chen2013theory, althammer2013quantitative, aqeel2016, fischer2018, geprags2020spin, becker2021electrical, hou2019spin}. The discovery of spin Seebeck effect (SSE) triggered various activities focused on thermal induction of magnon spin currents in a variety of complex magnetic phases to study the novel magnon spin transport phenomena \cite{uchida2008observation,rezende2016theory,cramer2018spin,mallick2019role}. Interestingly, in the recent years, observation of spin Hall magnetoresistance (SMR) has further opened up new avenues for the spin transport research \cite{aqeel2016,fischer2018,geprags2020spin,becker2021electrical,wagh2022probing}. The SMR can be described as follows; as spin Hall effect (SHE)-induced conduction-electron spin current in the HM layer approaches the MI$\vert$HM interface, if the spin current gets partially absorbed due to spin transfer torque (STT) then, it results in change in the HM resistivity \cite{nakayama2013spin,kato2020microscopic,zhang2019theory,reiss2021theory,chen2016theory}. Generally, resolving complex structures in canted-, spiral- and antiferro-magnets require use of advanced techniques like muon spin-relaxation, Lorentz transmission electron microscopy or spin-polarized neutron scattering \cite{wang2017,geprags2020spin,ganzhorn2016spin}. Recently, a novel approach of employing SMR as an electrical probe to examine exotic magnetic structures, has helped in understanding various magnetic phases and transformations \cite{aqeel2016,aqeel2015,becker2021electrical,hoogeboom2021magnetic,wang2017,guo2020nonlocal,chen2019spin,cheng2018antiferromagnet}. 

Functional materials with complex magnetic structures can provide new ways of explorations in spin transport devices. Typically, single phase magnetoelectric (ME) multiferroic materials are insulators and possess exotic magnetic structures (e.g. spiral, conical, etc) that can be controlled by both magnetic ($\bm{H}$) and electric ($\bm{E}$) field \cite{lu2019single, pullar2014multiferroic,rowley2016uniaxial}. Spin transport investigations on such ME multiferroic materials may open up a possibility of electric (voltage) control of the spin transport, a lucrative route to energy-efficient spintronic applications. Spiral magnetism that originates from strong competing ferro- and antiferro-magnetic interactions has been a popular route for exploring new multiferroic materials. Hexaferrite materials (M-type \ch{(Ba, Sr)(Fe, Sc, Mg)12O19}, Y-type \ch{(Ba, Sr)2Me2Fe12O22} and  Z-type \ch{(Ba, Sr)3Me2Fe24O41}) exhibit spiral magnetic phases stable over a wide range of temperatures spanning up to the room temperature (RT). This made hexaferrites a hot spot of research activities over the past decade \cite{kimura2012magnetoelectric, tokunaga2010multiferroic, lee2012heliconical, zhai2017giant, PhysRevB.97.134406,tang2015dielectric}. These materials are also known for their extensive usage as permanent magnets and their high potential in microwave applications. Notably, Z-type hexaferrites exhibit strong ME coupling near RT \cite{shin2022observation, wu2021memory, xu2019spin} making them a good candidate for ME applications. In recent studies, the role of dynamic ME coupling in high frequency (THz) applications (induction of electromagnons via activation of electric dipoles) has been emphasized \cite{chun2018electromagnon, kadlec2016electromagnon}. In the present work, we study spin transport in low-field, RT ME Z-type hexaferrite, \ch{Sr_3Co_2Fe_{24}O_{41}} (SCFO)$\vert$\ch{Pt} hybrid \cite{kitagawa2010,ueda2019insights,soda2011magnetic}. The motivation of our study is to unravel the influence of $\bm{H}$-tunable transverse conical magnetic phase in SCFO on the spin transport at SCFO$\vert$\ch{Pt} interface.

\begin{figure}
\includegraphics[width=0.48\textwidth,keepaspectratio]{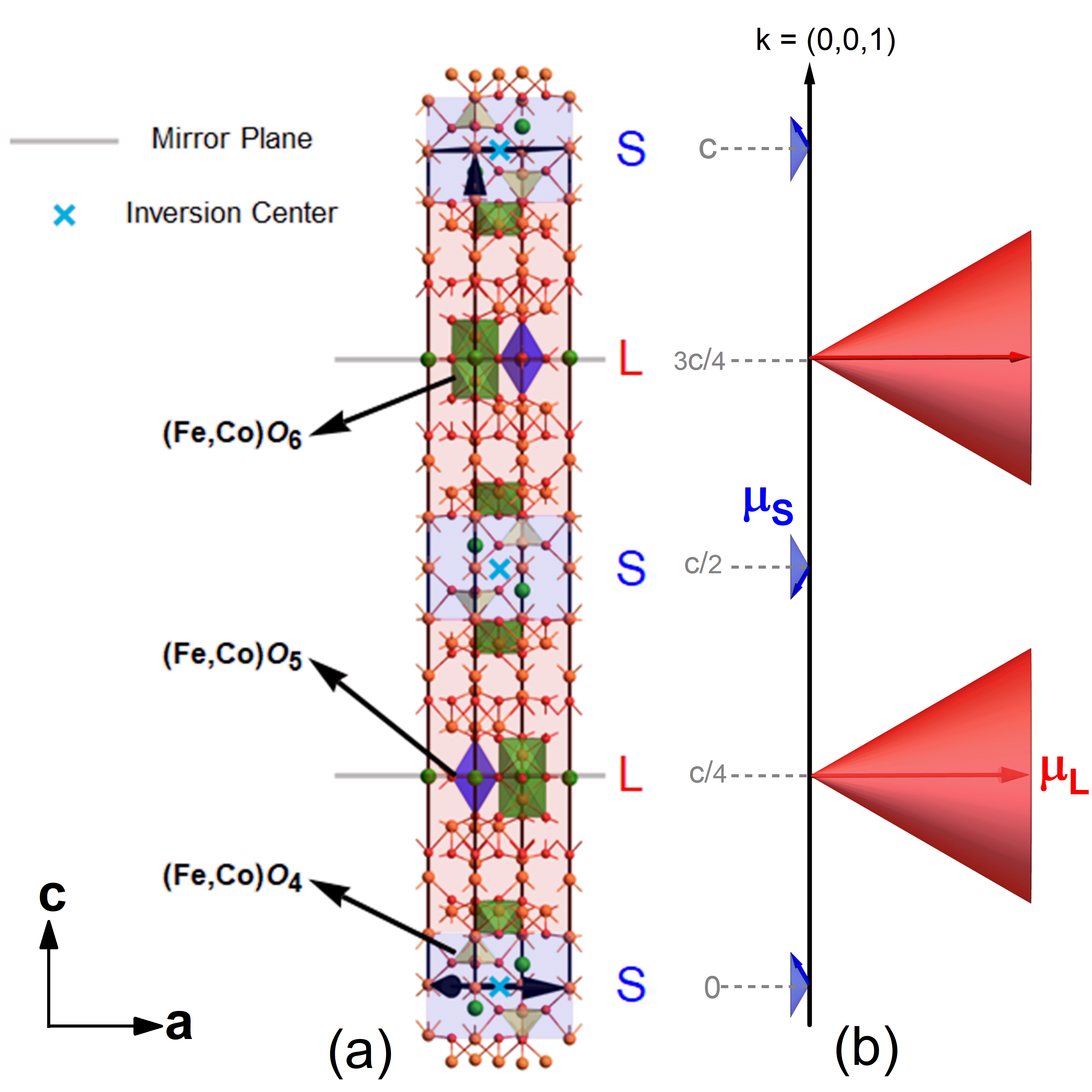}
\caption{\label{fig:1} A schematic of Z-type hexaferrite, \ch{Sr_3Co_2Fe_{24}O_{41}} (a) Crystal structure with alternate L and S magnetic blocks, and (b) Transverse conical magnetic structure at room temperature.}
\end{figure}

\ch{Co_2} Z-type hexaferrite SCFO exhibits hexagonal crystal structure with space group $P6_{3}/mmc$ and lattice parameters; $a = 5.8586$ {\AA} and $c = 51.9304$ {\AA}. The unit cell contains 30 transition-metal ions, \ch{Fe^{3+}} and \ch{Co^{2+}}, with octahedral, tetrahedral, and bi-pyramidal five-fold coordination with \ch{O^{2+}} ions \cite{nakajima2017observation}. The complex magnetism in SCFO is composed of alternating magnetic blocks (L and S) wherein magnetic moments align collinearly within each block \cite{chai2018magnetoelectricity,zhai2018room,burvsik2019hexaferrite}. A superexchange path inside the magnetic block causes anitiferromagnetic couplings along the path and manifests in ferrimagnetic interactions between the moments. Thus, the magnetic moment of the whole block can be represented by the net moment, $\mu_{\text{L}}$ = 39.8 $\mu_{\text{B}}$ ($\mu_{\text{S}}$ = 5.9 $\mu_{\text{B}}$) of the block L (S) \cite{utsumi2007superexchange,kimura2012magnetoelectric,ueda2019insights}. Superexchange interactions between the neighbouring blocks are antiferromagnetic in nature. The additional next-neighbour superexchange interactions play a crucial role in realizing non-collinear spiral structure in SCFO \cite{chun2018electromagnon}. Interestingly, single-ion anisotropy interactions compete with the superexchange interactions at the boundary between L- and S-blocks and, in turn, yield disparate magnetic structures at different temperatures \cite{chun2018electromagnon}. For instance, in SCFO paramagnetic phase transforms to $c$-axis-aligned ferrimagnetic phase at 690 K. Near 490 K, the ferrimagnetic phase aligns close to the basal plane ($c$-plane). Below 410 K, transverse conical phase is formed with a wave vector aligned parallel to the $c$-axis (See Fig. \ref{fig:1}(b)) \cite{chun2018electromagnon,nakajima2017observation}. In this phase, moments $\bm{\mu} _{\text{S}_1}$, $\bm{\mu} _{\text{L}_1}$, $\bm{\mu} _{\text{S}_2}$ and $\bm{\mu} _{\text{L}_2}$ are located along the $c$-axis at positions $0$, $c/4$, $c/2$ and $3c/4$, respectively \cite{ueda2019insights}. Notably, an inversion center and a mirror-plane lie at the center of each S-block and L-block, respectively (See Fig. \ref{fig:1}(a)) \cite{kimura2012magnetoelectric,kitagawa2010}. The symmetry considerations are generally known to prohibit Dzyaloshinskii-Moriya (DM) interactions between these magnetic blocks \cite{kimura2012magnetoelectric}. SCFO being the member of the ferroxplana family, its basal plane ($c$-plane) is an easy plane at RT, where a tiny $H$ is sufficient to rotate magnetic moments about c-axis and relatively large $H$ is needed to tilt the magnetic moments away from the $c$-plane \cite{ebnabbasi2012,takada2006}. This results in high permeability till the GHz regime, making SCFO a potential candidate for inductor cores at those frequencies. 

In this report, we present SSE and SMR measurements on polycrystalline SCFO$\vert$Pt hybrid. The SMR results are interpreted with the help of computation using a simple model.

\section{EXPERIMENTAL DETAILS}

\begin{figure}
\includegraphics[width=0.48\textwidth,keepaspectratio]{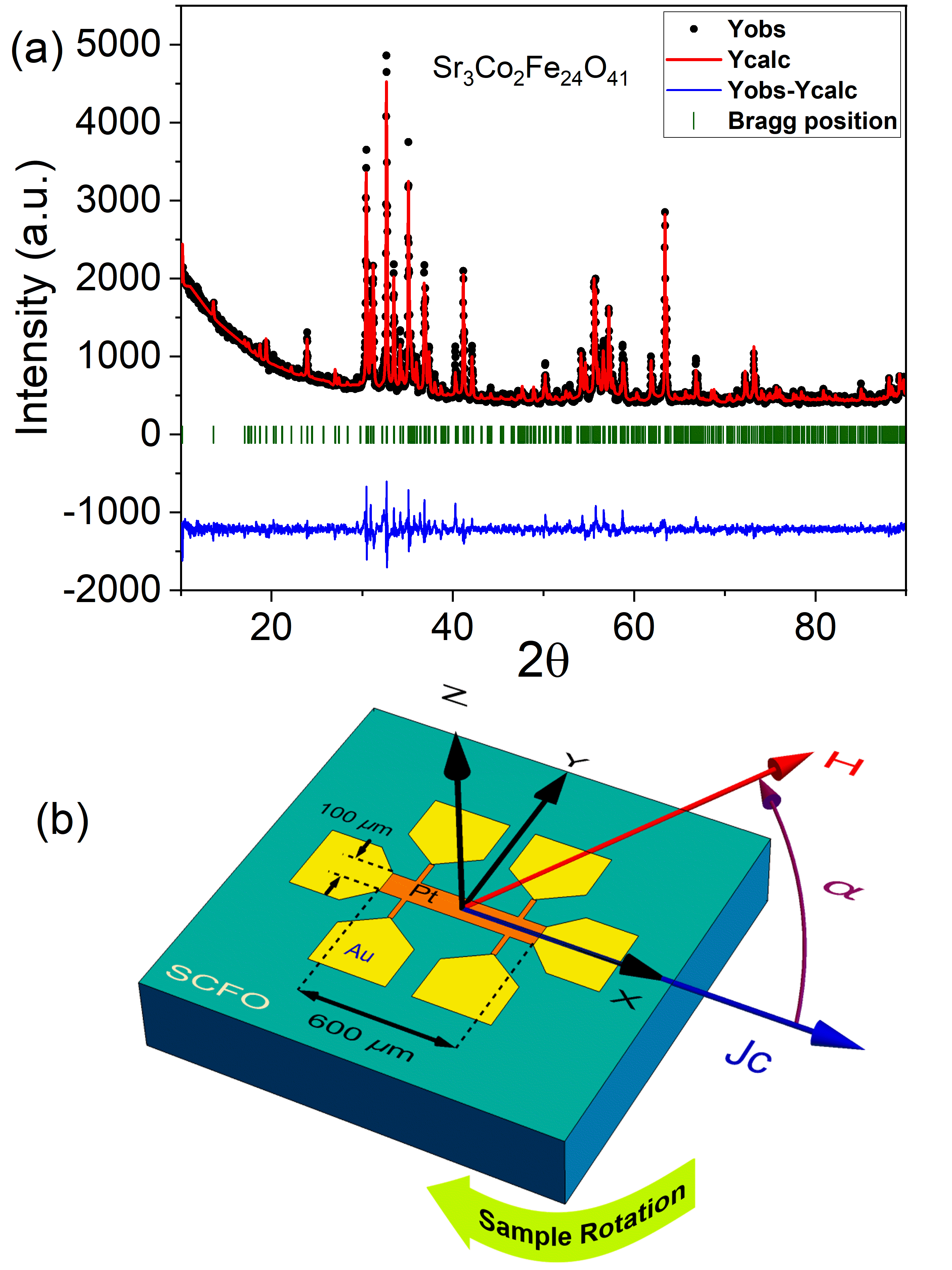}
\caption{\label{fig:2}(a): Powder XRD of  \ch{Sr_3Co_2Fe_{24}O_{41}}, (b) Schematic of the Hall-bar device.}
\end{figure}

\begin{figure*}
\begin{center}
\includegraphics[width=1\textwidth,keepaspectratio]{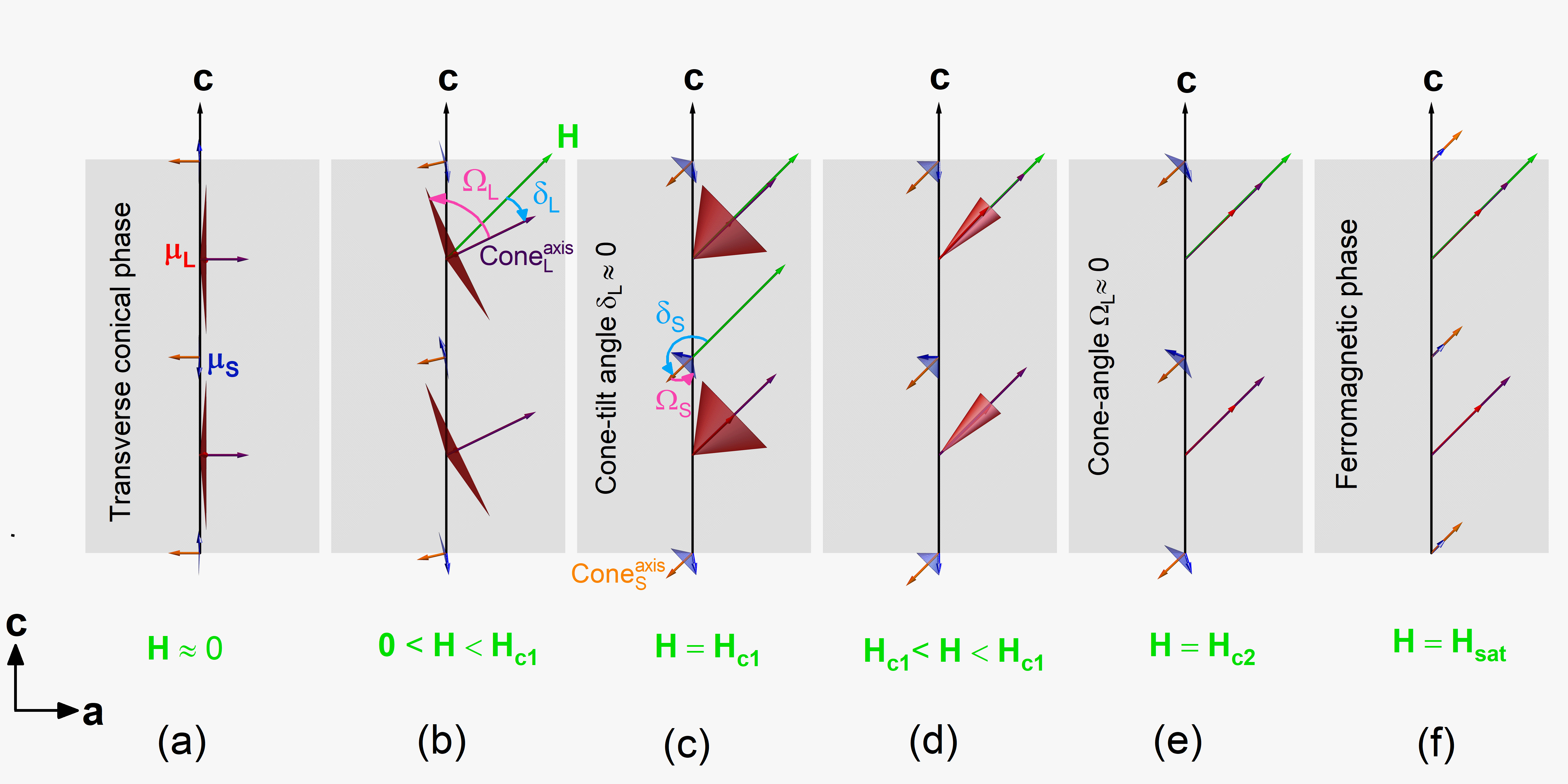}
\end{center}
\caption{\label{fig:3} Effect of slanted magnetic field, $\bm{H}$ on \ch{Sr_3Co_2Fe_{24}O_{41}} magnetic structure. (a) Transverse conical phase near zero field, (b) Low-field tilting of cone-axis, (c) Complete tilting of cone-axis ($Cone_{\text{L}}^{\text{axis}}$) at a critical field, $H_{\text{c1}}$, (d) High-field closing of cone-angle, $\Omega_{\text{L}}$, (e) Complete closing of cone-angle, $\Omega_{\text{L}}$ at a critical field, $H_{\text{c2}}$ and (f) Collinear ferromagnetic transformation at $H_{\text{sat}}$.}
\end{figure*} 

\begin{figure*}
\includegraphics[width=1\textwidth,keepaspectratio]{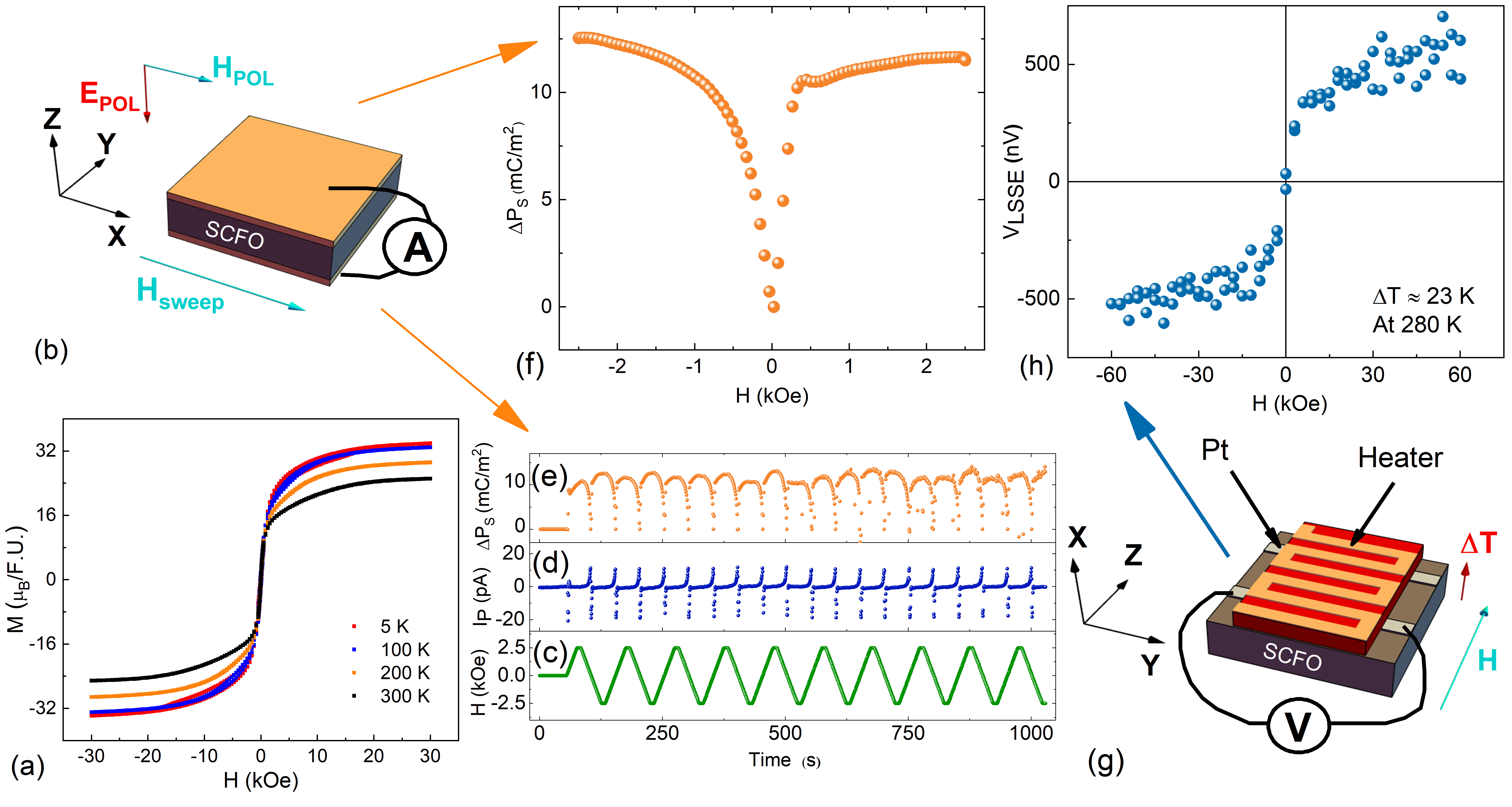}
\caption{\label{fig:4} (a) Isothermal $M$ Vs. $H$ curves recorded in \ch{Sr_3Co_2Fe_{24}O_{41}} at various temperatures. Magnetoelectric (ME) coupling studies: (b) schematic of ME device, (c) $H$-cycling with time, (d) ME current recorded during the $H$-cycling, (e) Estimated $\Delta P_{\text{S}}$ Vs. time, and (f) $H$-dependence of $\Delta P_{\text{S}}$. In longitudinal spin Seebeck effect (LSSE) measurements: (g) schematic of LSSE device and (h) $V_{\text{LSSE}}$ Vs. $H$ measured with applied thermal gradient of 23 K at base temperature of 280 K.} 
\end{figure*}

Powder samples of \ch{SrCO_3}, \ch{Co_3O_4} and \ch{Fe_2O_3} were mixed in stoichiometric ratio, well ground and cold pressed in to pellets of 15 mm diameter which were calcined at $1000$ C for 16 h in air \cite{arima2007ferroelectricity}. The pellets were pulverized and sintered again at $1200$ C for 16 h in air after pressing in to pellets of 15 mm diameter. Powder X-ray diffraction (XRD) (Rigaku SmartLab X-Ray Diffractometer) data of SCFO were analysed using Rietveld refinement method. The crystal structure was refined in the hexagonal space group $P6_{3}/mmc$ \cite{kitagawa2010}. Single phase was confirmed with the refined lattice parameters; $a = 5.8586$ {\AA} and $c = 51.9304$ {\AA} (Fig. \ref{fig:2}(a)). 
dc magnetization of a small piece of SCFO pellet was measured using vibrating sample magnetometer, PPMS, Quantum Design. In order to study the ME effect, a pellet was cut in a plate-like geometry (thickness = 0.5 $mm$ and area = 3 $mm^{2}$) and electrodes were silver painted on two opposing wide surfaces (Fig. \ref{fig:4}(b)). ME poling was carried out at 30 kOe and 160 V and subsequently, the magnetic field was swept at 100 Oe/s and the compensation current was measured using Keithley 6514A Electrometer.
For spin transport studies, the SCFO pellet was first cut by diamond cutter in rectangular dimension (4 mm x 3 mm) with thickness 1.5 mm. It was then polished (using BUEHLER MINIMET 1000 Grinder Polisher) with abrasive paper followed by polishing cloth and diamond paste with particle size of 0.25 $\mu$m. Finally, a \ch{Pt} film of thickness 5 nm was deposited at RT by electron-beam evaporation technique.

Hall-bar (600 $\mu$m long $\times$ 100 $\mu$m  wide) was patterned on the pellet using optical-lithography and Argon-ion etching technique. A thin layer of Au$\vert$Cr (100 nm$\vert$20 nm) was sputter-deposited as contact pads on the \ch{Pt} Hall-bar for spin transport measurements. 
 
In our SMR studies, we carried out two separate sets of measurements; (1) \textit{Measurement Set 1}- The sample was mounted on a rotating-probe, custom-built for the Lakeshore electromagnet (2 T). AC current of amplitude, $I_{\text{rms}}$ = 10/$\sqrt{2}$ $\mu$A (333 Hz) was passed using Keithley 6221 DC/AC Current Source and longitudinal signal was measured with Stanford Research SRS830 lock-in amplifier. The signal was measured at constant $H$ value, while the sample was rotated ($\alpha$-scan) as shown in the Fig. \ref{fig:2}(b). Angular dependence of SMR was measured at various $H$ values in the range, 50 Oe to 17 kOe. (2) \textit{Measurement Set 2}- The sample was mounted on Horizontal Rotator for PPMS 14 T, Quantum Design. AC current of amplitude, $I_{\text{rms}}$ = 500 nA (23.63 Hz) was applied and longitudinal signal was measured at constant in-plane $H$ levels in the range, 300 Oe to 100 kOe using PPMS 14 T. Here, the sample was rotated to attain $\alpha$-scan as shown in the Fig. \ref{fig:2}(b).

Longitudinal SSE measurements were carried out under high vacuum ($\approx 10^{-6}$ Torr) using a home-built probe for the PPMS 14 T, Quantum Design. Out of plane thermal gradient ($\Delta T$) was applied and the longitudinal SSE signal was measured using Keithley 2182A Nanovoltmeter. Details of the device geometry are discussed in Section \ref{MH_ME_SSE}.

\section{H-induced tilting and closing of cone in transverse conical phase}

At RT, in the absence of external field, SCFO exhibits transverse conical phase with wave vector along $c$-axis, where $\bm{\mu}_{\text{L}_{(1,2)}}$ magnetic moments are in $c$-plane and $\bm{\mu} _{\text{S}_{(1,2)}}$ magnetic moments are inclined out of $c$-plane (Fig. \ref{fig:3}(a)). Thus, the cone-axes lie in the easy plane (basal plane) of the unit cell in SCFO. We define cone-angle, $\Omega_{\text{L}}$  ($\Omega_{\text{S}}$) to be an angle subtended between a cone-axis, $Cone_{\text{L}}^{\text{axis}}$ ($Cone_{\text{S}}^{\text{axis}}$) and corresponding magnetic moment, $\bm{\mu} _{\text{L}_{(1,2)}}$ ($\bm{\mu} _{\text{S}_{(1,2)}}$) (See Fig. \ref{fig:3}(b) and (c)). Also, a cone-tilt angle, $\delta_{\text{L}}$  ($\delta_{\text{S}}$) is defined as an angle between the cone-axis, $Cone_{\text{L}}^{\text{axis}}$ ($Cone_{\text{S}}^{\text{axis}}$) and $\bm{H}$ (Fig. \ref{fig:3}(b) and (c)). When a small $\bm{H}$ is applied in a slanted direction away from the basal plane, it predominantly induces tilting of the $Cone_{\text{L}}^{\text{axis}}$ ($Cone_{\text{S}}^{\text{axis}}$) towards the direction of $\bm{H}$ ($-\bm{H}$) and results in decrease (increase) in the value of $\delta_{\text{L}}$  ($\delta_{\text{S}}$) (Fig. \ref{fig:3}(b)). The process of tilting of cone would continue until $\bm{H}$ reaches a critical field, $\bm{H}_{\text{c1}}$ where $Cone_{\text{L}}^{\text{axis}}$ aligns along $\bm{H}$ (i.e. $\delta_{\text{L}} = 0$) (Fig. \ref{fig:3}(c)). Further increase in $\bm{H}$ exclusively results in gradual closing of the cone (i.e. decrease in $\Omega_{\text{L}}$) (Fig. \ref{fig:3}(d)). At still higher field near another critical field $\bm{H} = \bm{H}_{\text{c2}}$, the cone closes completely (i.e. $\Omega_{\text{L}} = 0$) (Fig. \ref{fig:3}(e)). At saturation field ($\bm{H} = \bm{H}_{sat}$), both $\delta_{\text{S}}$ and $\Omega_{\text{S}}$ tend to zero and this results in collinear ferromagnetic ordering (Fig. \ref{fig:3}(f)).

To summarize, an application of $\bm{H}$ along an arbitrary direction in a single unit cell yields certain amount of tilting or/and closing of the cones. However, if $\bm{H}$ is applied across a polycrystalline sample, then the amount of the tilting or/and closing of the cones would be different for distinct randomly-oriented grains. This is due to the fact that the orientation of the $\bm{H}$ with respect to each randomly oriented grain is different.

\section{EXPERIMENTAL RESULTS}

\subsection{\label{MH_ME_SSE} Magnetization, electric polarization and spin Seebeck effect}

\begin{figure*}
\includegraphics[width=1\textwidth,keepaspectratio]{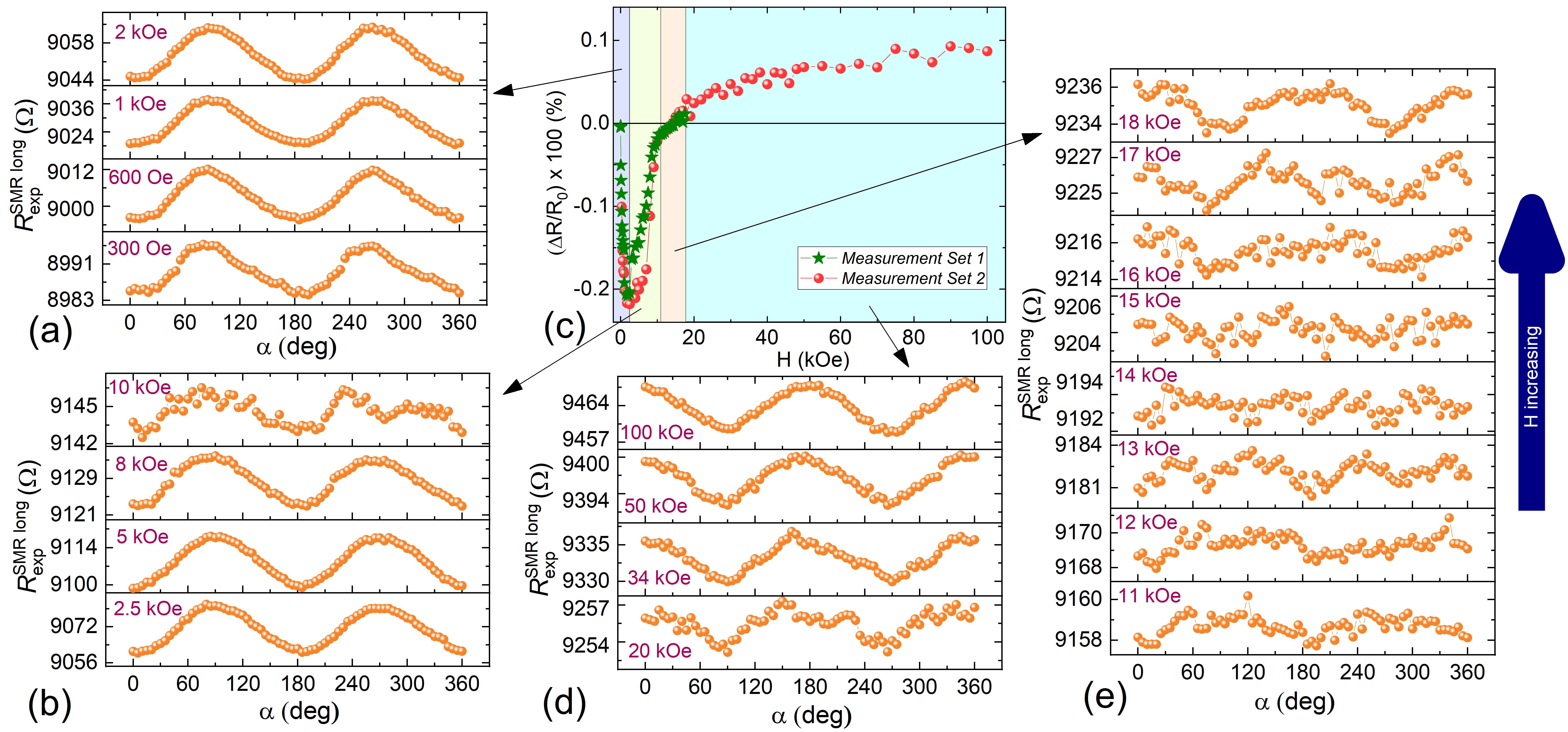}
\caption{\label{fig:5}(a), (b), (d) and (e) Longitudinal SMR signal ($R_{\text{exp}}^{\text{SMR long}}$), measured during $\alpha$-scan at various $H$-values in the range, 300 Oe to 100 kOe (\textit{Measurement Set 2} data). (c) Normalized amplitude of $R_{\text{exp}}^{\text{SMR long}}$-modulation, $(\Delta R / R_{0}) \times 100 \%$ plotted as a function of $H$ for \textit{Measurement Set 2} data and compared with \textit{Measurement Set 1} data (Refer to Fig. \ref{fig:S1} for the details).}
\end{figure*}

Figure \ref{fig:4}(a) shows the plot of isothermal dc magnetization ($M$) Vs. $H$ data for a polycrystalline SCFO bulk sample measured at various temperatures viz. 5, 100, 200 and 300 K. It is observed that $M$ increases rapidly with $H$ at low $H$ values and, then at moderate rates over a certain $H$-range. At higher fields, it approaches a saturation level very gradually. At 300 K and 2.4 kOe, the $M$ rises to $\approx 61 \%$ of the apparent saturation magnetization, $M_{\text{sat}}$. At lower temperatures, increase in the $M_{\text{sat}}$ is expected. The value of $M_{\text{sat}}$ at 5 K is observed to be $\approx$ 33.9 $\mu _{\text{B}}/\text{F.U.}$. Besides, a small hysteresis is seen in $M$ Vs. $H$ data in the interval between 1 and 20 kOe at 5 and 100 K. These features are consistent with those observed in the literature although the reason is unclear \cite{kitagawa2010}. A sharp rise in $M$ at low field could be due to prominent decrease in the cone-tilt angle ($\delta _{\text{L}}$) whereas, a gradual rise at high fields may be due to closing of the cone-angles, $\Omega _{\text{L}}$ and $\Omega _{\text{S}}$.

For ME characterization of SCFO, the sample was poled with 160 V (along Z-axis) under applied $H$ of 30 kOe (along X-axis) (See Fig. \ref{fig:4}(b)). Then, the $H$ was decreased to -2.5 kOe and the poling electric field was removed. Subsequently, $H$ was scanned between -2.5 and 2.5 kOe and ME current was measured (Fig. \ref{fig:4}(c) and (d)). Estimated change in spontaneous polarization ($\Delta P_{\text{S}}$) with time (or $H$-cycling) is shown in Fig. \ref{fig:4}(e). The $H$-variation of $\Delta P_{\text{S}}$ is plotted in Fig. \ref{fig:4}(f). $\Delta P_{\text{S}}$ is nearly zero at zero field and with increase (or decrease) in $H$, it increases sharply to positive values. The sign retention of $\Delta P_{\text{S}}$ across zero field in $H$-scan has been recently attributed to a new dominant mechanism of polarization, namely $p$$\text{-}$$d$ hybridization \cite{ueda2019insights}. Besides, classical spin-current model contributes a relatively moderate $\Delta P_{\text{S}}$ (with antisymmetric variation) which makes the observed $\Delta  P_{\text{S}}$ slightly asymmetric about $H$= 0. The tilting of the cones at low-field is responsible for the asymmetry.  

In the longitudinal SSE study at 280 K, we applied out-of-plane thermal gradient ($\Delta T \approx 23$ K) across the device using gold heater patterned on sapphire substrate (Fig. \ref{fig:4}(g)) and applied $H$ in-plane and perpendicular to the platinum bar. Longitudinal SSE signal ($V_{\text{LSSE}}$) Vs. $H$ plot (Fig. \ref{fig:4}(h)) resembles $M$ Vs. $H$ curve (rapid increase at low-$H$ and gradual increase at high-$H$) and gives $V_{\text{LSSE}} \approx 500$ nV at 60 kOe. The observed rapid change in $V_{\text{LSSE}}$ below 3 kOe indicates that the thermally-induced pure magnon spin currents are influenced by the tilting of the cones in low field regime. For instance, $V_{\text{LSSE}}$ rises to $\approx 229$ nV at 3 kOe.

\subsection{\label{exp_smr}Spin Hall magnetoresistance}

In SCFO $\vert$\ch{Pt} device, when the charge current is passed through a \ch{Pt} Hall-bar along X-direction (See Fig. \ref{fig:2}(b)), the SHE due to high spin-orbit coupling generates transverse spin current which, in turn, results in spin accumulation ($\bm{\mu} _{\text{spin}}$) at the SCFO$\vert$\ch{Pt} interface. The relative angle between the corresponding spin polarization ($\bm{S}$) in Pt and $\bm{M}$ in SCFO dictates the amount of absorption of the spin current at the interface via STT. This leads to a net spin current in the \ch{Pt}-layer and results in transverse charge current via inverse SHE. As a consequence, the rotation of $\bm{H}$ manifests in modulation of \ch{Pt} resistance.

In \textit{Measurement Set 2}, we applied charge current ($I_{\text{rms}}$ = 500 $n \text{A}$) along X-direction in the \ch{Pt} Hall-bar and rotated $\bm{H}$ in-plane ($\alpha$-scan) (Fig. \ref{fig:2}(b)) and measured longitudinal SMR (i.e. $R_{\text{exp}}^{\text{SMR long}}$ Vs. $\alpha$) at various constant amplitudes of $\bm{H}$ (300 Oe to 100 kOe) as plotted in Fig. \ref{fig:5}(a), (b), (d) and (e). The procedure adopted to separate out thermal drift from the raw $\alpha$-scan data is elaborated in Section \ref{appendix_set2} in the Appendix. The amplitude of $R_{\text{exp}}^{\text{SMR long}}$-modulation (i.e. $\Delta R$) was extracted for each $\alpha$-scan and the percentage SMR ($(\Delta R / R_{0}) \times 100 \% $) was estimated and plotted as a function of $H$ (Fig. \ref{fig:5}(c)). The extracted $(\Delta R / R_{0}) \times 100 \% $ Vs $H$ data in \textit{Measurement Set 1} (the corresponding $\alpha$-scan data are presented in Section \ref{appendix_set1} in the Appendix) are plotted together along with \textit{Measurement Set 2} data for comparison (Fig. \ref{fig:5}(c)). Both sets of $(\Delta R / R_{0}) \times 100 \% $ Vs $H$ curves match closely, which indicates the consistency of the SMR measurements.

It is discernible in Fig. \ref{fig:5}(a) and (b) that at low fields (300 Oe to 2 kOe), the angular dependence of $R_{\text{exp}}^{\text{SMR long}}$ Vs. $\alpha$ shows sinusoidal signal with $180$ deg periodicity, having peaks at $90$ deg and $270$ deg. This implies that the observed SMR is negative. As $H$ increases ($H <$ 2.5 kOe), the slope of $(\Delta R / R_{0}) \times 100 \%$ Vs. $H$ curve is negative (See light blue region in Fig. \ref{fig:5}(c)). Around 2.5 kOe, $(\Delta R / R_{0}) \times 100 \%$ attains the lowest value of $-0.21 \%$. In the field range; 2.5 kOe $\leq H <$ 10.5 kOe, the amplitude of the negative SMR modulation decreases with increase in $H$. In this regime, the $(\Delta R / R_{0}) \times 100 \%$ Vs. $H$ curve (light green region in Fig. \ref{fig:5}(c)) exhibits a positive slope. With further increase in $H$ (10.5 kOe $\leq H <$ 18 kOe), the slope of $(\Delta R / R_{0}) \times 100 \%$ Vs. $H$ curve is drastically reduced. Subsequently, a cross-over from negative to positive value is seen (light orange region in Fig. \ref{fig:5}(c)). In order to probe the evolution of this cross-over, a few $\alpha$-scans (at $H$ = 11, 12, 13, 14, 15, 16, 17 and 18 kOe) are recorded and plotted in Fig. \ref{fig:5}(e). After the cross-over region ($H \geq$ 18 kOe), there is a gradual change in the positive slope of $(\Delta R / R_{0}) \times 100 \%$ Vs. $H$ curve with increase in field (See light cyan region in Fig. \ref{fig:5}(c)). Here, $R_{\text{exp}}^{\text{SMR long}}$ Vs. $\alpha$ shows positive SMR modulation with peaks at $0$ deg and $180$ deg as shown in Fig. \ref{fig:5}(d)). The amplitude of positive SMR increases with $H$.

\section{COMPUTATION OF SPIN HALL MAGNETORESISTANCE}

\subsection{Method}

\begin{figure}
\includegraphics[width=0.48\textwidth,keepaspectratio]{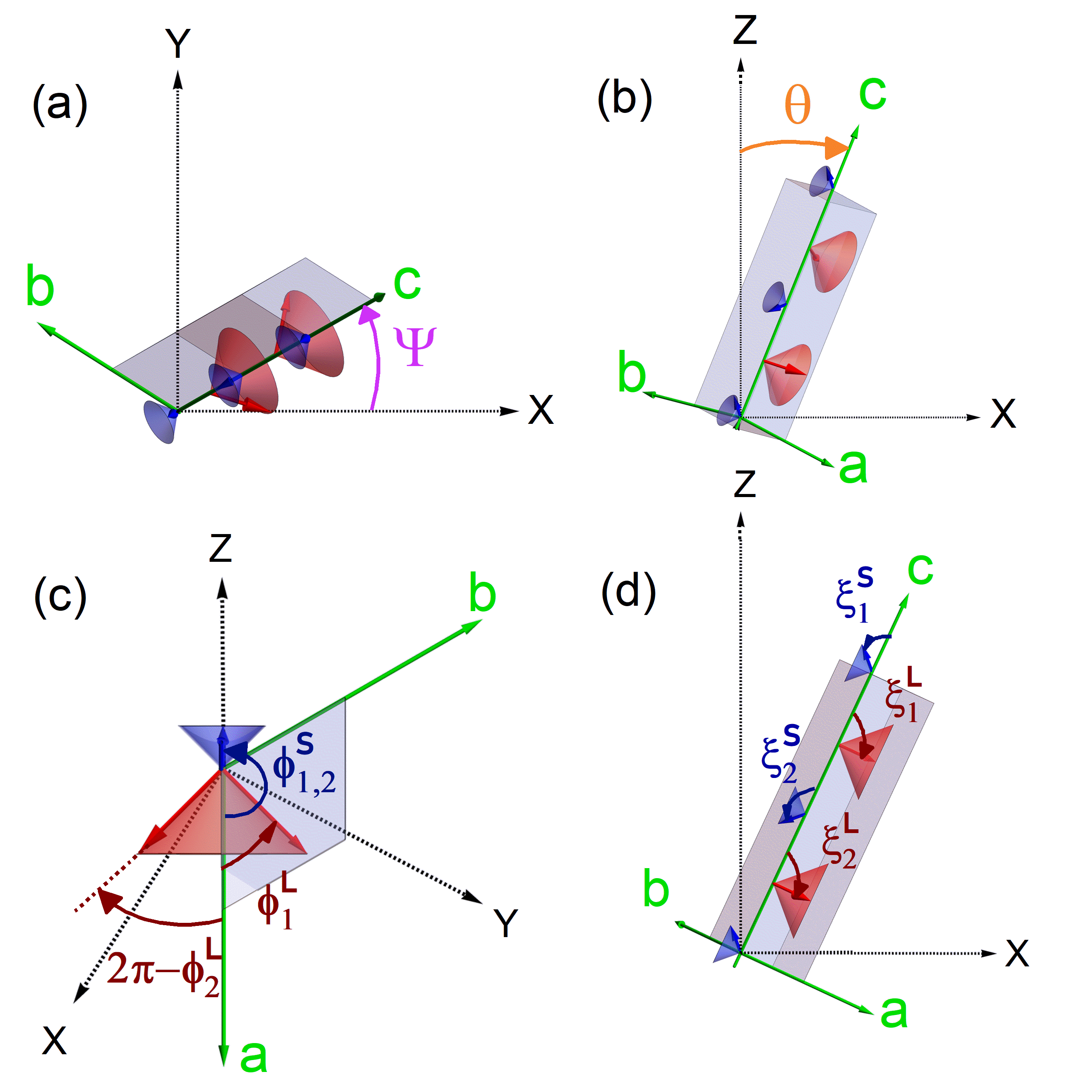}
\caption{\label{fig:6} Representation of a tilted unit cell inside a polycrystalline sample (XYZ reference frame). Angles $\theta$ and $\psi$ ( (b) and (a) ) represent orientation of the unit cell in the sample while $\xi$ and $\phi$ ( (d) and (c) ) denote local angles that magnetic moments make with $c$ and $a$-axis.}
\end{figure}

We measured longitudinal dc magnetization and SMR in polycrystalline samples under applied magnetic fields. The measured signals consist of the contributions from various randomly oriented grains. In order to model this scenario, we first consider a unit cell wherein, the $c$-axis ($a$-axis) is aligned along the Z-axis (X-axis) of the sample. To generate random orientations of the grain (or the unit cell), we rotate the $c$-axis ($a$-axis) of the unit cell by the angle, $\theta$ ($\psi$) about Y-axis (Z-axis) of the sample (See Fig. \ref{fig:4}(a) and (b)). In our SMR studies, initially the $\bm{H}$ was applied along the X-axis of the polycrystalline sample and, then $\bm{H}$ was rotated by an angle ($\alpha$), about the Z-axis (See Fig. \ref{fig:2}(b)).

The equilibrium magnetization inside a magnetic unit cell of SCFO at a given magnetic field can be estimated by free energy minimization. A magnetic structure is composed of two types of magnetic blocks, L and S, with their representative magnetic moments, $\bm{\mu} _{\text{L}_{(1,2)}}$ and $\bm{\mu} _{\text{S}_{(1,2)}}$, respectively. Further, $\bm{\mu} _{\text{L}_{(1,2)}}$ and $\bm{\mu} _{\text{S}_{(1,2)}}$ can be expressed by Eq.\ref{eq:ml} and Eq.\ref{eq:ms}, respectively. For a particular orientation of the unit cell (unique $\theta$ and $\psi$), moments, $\bm{\mu} _{\text{L}_{(1,2)}}$ ($\bm{\mu} _{\text{S}_{(1,2)}}$), make local angles $\xi _{(1,2)}^{\text{L}}$ ($\xi _{(1,2)}^{\text{S}}$) with $c$-axis and $\phi _{(1,2)}^{\text{L}}$ ($\phi _{(1,2)}^{\text{S}}$) with $a$-axis of the unit cell (Fig. \ref{fig:4}(d) and (c)).

The total magnetic free energy, $E$ for a particular orientation of the unit cell ($\theta$ and $\psi$ are unique) can be expressed as in Eq.\ref{freeenergy} \cite{talbayev2008detection,chun2018electromagnon}. The first three terms in the equation represent superexchange interactions between $\bm{\mu}_{\text{L}_{1}}$\text{$\leftrightarrow$}$\bm{\mu}_{\text{L}_{2}}$, $\bm{\mu}_{\text{S}_{1}}$\text{$\leftrightarrow$}$\bm{\mu}_{\text{S}_{2}}$ and  $\bm{\mu}_{\text{L}_{(1,2)}}$\text{$\leftrightarrow$}$\bm{\mu}_{\text{S}_{(1,2)}}$ with the constants $J_{\text{LL}}$, $J_{\text{SS}}$ and $J_{\text{LS}}$, respectively. The forth and the fifth terms, consisting anisotropy constants $D_{\text{L}}$ and $D_{\text{S}}$ for $\bm{\mu}_{\text{L}}$ and $\bm{\mu}_{\text{S}}$, represent single-ion anisotropy. The sixth term describes DM interaction characterized by a vector, $\bm{D}_{\text{DM}_{ij}}$ associated with a pair of spin moments, $\bm{\mu}_{\text{L}_{\text{i}}}$ and $\bm{\mu}_{\text{Sj}}$. The last two terms represent the Zeeman energy due to applied magnetic field $\bm{H} = H \cos \left(\alpha\right) \hat{\bm{x}} + H \sin \left(\alpha\right) \hat{\bm{y}}$. Typically, due to symmetry considerations in hexaferrites, DM interactions are considered to be absent \cite{talbayev2008detection,chun2018electromagnon}. However, a recent report \cite{chun2018electromagnon} argued that in order to explain their electromagnon resonance spectra of Z-type hexaferrite, inclusion of DM interactions of moderate strength was required. Notably, the strength of DM interactions was taken as low as 8$\%$ of the strength of the respective major superexchange interaction. It was also considered proportional to the static electric polarization which, in turn, depended on $H$. In order to keep our model simple, we have neglected DM interaction terms in Eq.\ref{freeenergy}. For computation, we considered the spin moments to be dimensionless ($\mu_{\text{L}}$ = 39.8 and $\mu_{\text{S}}$ = 5.9) and used the values; $J_{\text{LL}} = 4.42$ $\mu$eV, $J_{\text{SS}} = 241.9$ $\mu$eV, $J_{\text{LS}} = 58.97$ $\mu$eV, $D_{\text{L}} = 0.014$ neV and $D_{\text{S}} = -0.69$ neV. For comparison, we list here the values of exchange constants for another Z-type hexaferrite, \ch{Ba_{0.5}Sr_{2.5}Co2Fe24O41} reported elsewhere \cite{chun2018electromagnon}; $J_{\text{LL}} = 4.69$ $\mu$eV, $J_{\text{SS}} = 271.82$ $\mu$eV and $J_{\text{LS}} = 58.98$ $\mu$eV. The aniotropy constants used in our calculations were small in magnitudes but found essential for the simulation of the SMR-behavior. In our studies, $\bm{H}$ was either scanned along X-axis or rotated in XY-plane ($\alpha$-scan) in the sample. To estimate equilibrium magnetization in the unit cell, we minimize total free energy, $E\left(\xi _1^{\text{L}},\xi _2^{\text{L}},\xi _1^{\text{S}},\xi _2^{\text{S}},\phi _1^{\text{L}},\phi _2^{\text{L}},\phi _1^{\text{S}},\phi _2^{\text{S}}\right)$ with respect to $\xi _1^{\text{L}}$, $\xi _2^{\text{L}}$, $\xi _1^{\text{S}}$, $\xi _2^{\text{S}}$, $\phi _1^{\text{L}}$, $\phi _2^{\text{L}}$, $\phi _1^{\text{S}}$ and $\phi _2^{\text{S}}$.

\begin{widetext}

\begin{subequations}
\label{eq:ml1ms1}

\begin{equation}
\bm{\mu}_{\text{L}_{(1,2)}} =\mu_{\text{L}} 
\begin{pmatrix}
\left(\sin \theta  \cos \psi  \cos \xi _{(1,2)}^{\text{L}}+\sin \xi _{(1,2)}^{\text{L}} (\cos \theta  \cos \psi  \cos \phi _{(1,2)}^{\text{L}}-\sin \psi  \sin \phi _{(1,2)}^{\text{L}})\right) \hat{\bm{x}}\\
\left(\sin \theta  \sin \psi  \cos \xi _{(1,2)}^{\text{L}}+\sin \xi _{(1,2)}^{\text{L }} (\cos \theta  \sin \psi  \cos \phi _{(1,2)}^{\text{L }}+\cos \psi  \sin \phi _{(1,2)}^{\text{L }})\right)\hat{\bm{y}}\\
\left(\cos \theta  \cos \xi _{(1,2)}^{\text{L }}-\sin \theta  \sin \xi _{(1,2)}^{\text{L }} \cos \phi _{(1,2)}^{\text{L }}\right)\hat{\bm{z}}
\end{pmatrix}
\label{eq:ml}
\end{equation}

\begin{equation}
\bm{\mu} _{\text{S}_{(1,2)}} =\mu_{\text{S}} 
\begin{pmatrix}
\left(\sin \theta  \cos \psi  \cos \xi _{(1,2)}^{\text{S}}+\sin \xi _{(1,2)}^{\text{S}}(\cos \theta  \cos \psi  \cos \phi _{(1,2)}^{\text{S}}-\sin \psi  \sin \phi _{(1,2)}^{\text{S}})\right)\hat{\bm{x}}\\
\left(\sin \theta  \sin \psi  \cos \xi _{(1,2)}^{\text{S}}+\sin \xi _{(1,2)}^{\text{S}} (\cos \theta  \sin \psi  \cos \phi _{(1,2)}^{\text{S}}+\cos \psi  \sin \phi _{(1,2)}^{\text{S}})\right)\hat{\bm{y}}\\
\left(\cos \theta  \cos \xi _{(1,2)}^{\text{S}}-\sin \theta  \sin \xi _{(1,2)}^{\text{S}} \cos \phi _{(1,2)}^{\text{S}}\right)\hat{\bm{z}}
\end{pmatrix}
\label{eq:ms}
\end{equation}
\end{subequations}

\begin{eqnarray}
E\left(\xi_1^{\text{L}},\xi_2^{\text{L}},\xi_1^{\text{S}},\xi_2^{\text{S}},\phi _1^{\text{L}},\phi _2^{\text{L}},\phi _1^{\text{S}},\phi _2^{\text{S}}\right)=2J_{\text{LL}}\left(\bm{\mu}_{\text{L}_{1}}.\bm{\mu}_{\text{L}_{2}}\right)+2J_{\text{SS}}\left(\bm{\mu}_{\text{S}_{1}}.\bm{\mu}_{\text{S}_{2}}\right)+J_{\text{LS}}\sum _{\text{i,j}=1,2} \left(\bm{\mu}_{\text{L}_{\text{i}}}.\bm{\mu}_{\text{S}_{\text{j}}}\right)\nonumber\\
+D_\text{L}\sum _{\text{i}=1,2} \left(\bm{\mu}_{\text{L}_{\text{i}}}.\bm{c}\right)^2+D_\text{S}\sum _{\text{i}=1,2} \left(\bm{\mu}_{\text{S}_{\text{i}}}.\bm{c}\right)^2+\sum _{\text{i,j}=1,2} \bm{D}_{\text{DM}_{\text{ij}}}.\left(\bm{\mu}_{\text{L}_{\text{i}}}\times \bm{\mu}_{\text{S}_{\text{j}}}\right)-\underset{\text{i}=1,2}{\sum}\bm{H}.\left(\bm{\mu}_{\text{L}_{\text{i}}}+\bm{\mu}_{\text{S}_{\text{j}}}\right)
\label{freeenergy}
\end{eqnarray}

\begin{eqnarray}
\begin{aligned}
R_{\text{sim}}^{\text{SMR long}} =\frac{1}{2 \pi^{2}} \int _{\theta =0}^{\pi}\int _{\psi =0}^{2 \pi}R_0+\Delta R_{1} &((\left(\bm{\mu} _{\text{L}_{1}}\left(\theta ,\psi\right).\hat{\bm{x}}\right)^{2}+\left(\bm{\mu} _{\text{L}_{2}}\left(\theta ,\psi\right).\hat{\bm{x}}\right)^{2}\\
&+\left(\bm{\mu} _{\text{S}_{1}}\left(\theta ,\psi\right).\hat{\bm{x}}\right)^{2}+\left(\bm{\mu} _{\text{S}_{2}}\left(\theta ,\psi\right).\hat{\bm{x}}\right)^{2})/ 4) d\psi d\theta
\end{aligned}
\label{eq:A1}
\end{eqnarray}

\end{widetext}

\begin{figure*}
\includegraphics[width=1\textwidth,keepaspectratio]{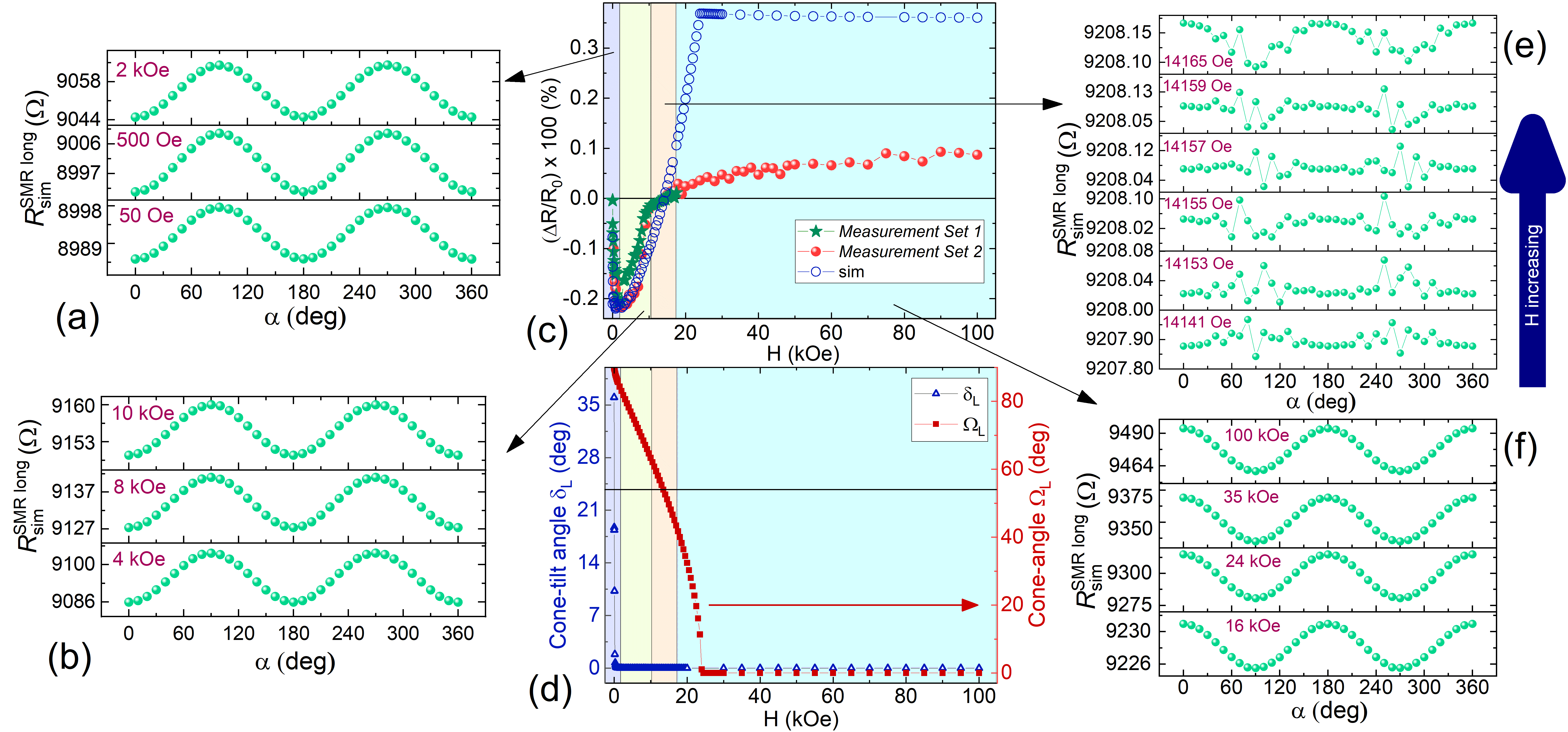}
\caption{\label{fig:7}(a), (b), (e) and (f) Simulated longitudinal SMR ($R_{\text{sim}}^{\text{SMR long}}$) for $\alpha$-scan at various $H$ values. (c) Normalized simulated SMR amplitude, $(\Delta R / R_{0}) \times 100 \%$ plotted as a function of $H$, and compared with experimental data (\textit{Measurement Set 1} and \textit{Measurement Set 2}), (d) simulated $H$-evolution of cone-tilting angle and cone-closing angle for $\bm{\mu} _{\text{L}_{(1,2)}}$. }
\end{figure*}

For SMR studies, $R_{\text{exp}}^{\text{SMR long}}$ was measured while rotating $\bm{H}$ in z-plane (See Fig. \ref{fig:2}(b)). In order to compute SMR for polycrystalline SCFO$\vert$Pt hybrid, we use simple model (described in Eq.\ref{freeenergy} and \ref{eq:A1}) based on the theory of SMR in collinear magnets. The relative scales of spin-relaxation length in \ch{Pt} and wavelength of the conical structure in SCFO favored the use of this theory. Aqeel \textit{et al}. \cite{aqeel2016} used this approach to study helical and longitudinal conical structures in \ch{Cu_{2}OSeO_{3}}. In our method, $R_{\text{sim}}^{\text{SMR long}}$ for a particular angle ($\alpha$) is computed using Eq.\ref{eq:A1}. The integrand in the Eq.\ref{eq:A1} can be computed for a particular orientation of the unit cell with the knowledge of X-projections of all the four magnetic moments ($\bm{\mu} _{\text{S}_{1}}$, $\bm{\mu} _{\text{S}_{2}}$, $\bm{\mu} _{\text{L}_{1}}$ and $\bm{\mu} _{\text{L}_{2}}$). Finally, $R_{\text{sim}}^{\text{SMR long}}$ is obtained by averaging the expression in the integrand over all random orientations of the cell i.e. by integrating it over the full range of $\psi$ and $\theta$.

\subsection{Computation Results}

In this section, we compare the computed results of SMR with experimentally measured SMR (\textit{Measurement Set 1} and \textit{Measurement Set 2}) and further, analyse the results in relation to $H$-evolution of the cone-tilt angle and the cone-angle. Normalized SMR amplitude $(\Delta R / R_{0}) \times 100 \%$ vs. $H$ curves calculated from experimental \textit{Measurement Set 1} and \textit{Measurement Set 2} are plotted together in four different $H$-regimes (as discussed in the Section \ref{exp_smr}) and compared with simulated $(\Delta R / R_{0}) \times 100 \%$ vs. $H$ (marked by blue circles) in Fig. \ref{fig:7}(c). These regimes are denoted by shaded regions with four distinct colors. At different $H$-values within each $H$-regime, SMR Vs. $\alpha$ curves have been simulated and are illustrated in Fig. \ref{fig:7}(a), (b), (e) and (f), respectively. Arrows indicate the mapping of the simulated curve and the corresponding $H$-regime. SMR at each $H$-value, is determined by free energy minimization and calculation of $\text{R}_{\text{sim}}^{\text{SMR long}}$ using Eq.\ref{freeenergy} and Eq.\ref{eq:A1}, respectively. We calculated the normalized SMR amplitude ($(\Delta R / R_{0}) \times 100 \%$) from the simulated $\alpha$-scans (Fig. \ref{fig:7}(a), (b), (e) and (f)) and plotted as a function of $H$ in Fig. \ref{fig:7}(c). Besides, with the help of computed equilibrium spin configuration ($\bm{\mu}_{\text{L}_{(1,2)}}$ spin moments) at different $H$ values, we estimated average values of cone-tilt angle ($\delta_{\text{L}}^{\text{avg}}$) and cone-angle ($\Omega_{\text{L}}^{\text{avg}}$). The $\delta_{\text{L}}^{\text{avg}}$ and $\delta_{\text{L}}^{\text{avg}}$ estimated for $\bm{\mu}_{\text{L}_{(1,2)}}$ Vs. $H$ are plotted in Fig. \ref{fig:7}(d).

In the low $H$ regime ($H < 2.5$ kOe, represented by light blue shaded region in Fig. \ref{fig:7}(c) and (d)), computed $\text{R}_{\text{sim}}^{\text{SMR long}}$ Vs. $\alpha$ curves illustrate negative SMR with peaks at $90$ deg and  $270$ deg (See Fig. \ref{fig:7} (a)). The negative slope of the computed $(\Delta R / R_{0}) \times 100 \%$ Vs. $H$ curve in this $H$-regime (Fig. \ref{fig:7}(c)) is attributed to prominent tilting of cones below 2.5 kOe (Fig. \ref{fig:7}(d)). Here, $\delta_{\text{L}}^{\text{avg}}$ reduces from about 35 deg to $\approx 0$ deg by 2.5 kOe.

Simulated $\text{R}_{\text{sim}}^{\text{SMR long}}$ Vs. $\alpha$ signal in second $H$-regime ($2.5 \leq H < 10.5$ kOe; light green shaded region in Fig. \ref{fig:7}(c) and (d)) also showed similar negative SMR behaviour (See Fig. \ref{fig:7} (b)). The slope of the estimated $(\Delta R / R_{0}) \times 100 \%$ Vs. $H$ curve changes from negative to positive value and a dip around 2.5 kOe ($(\Delta R / R_{0}) \times 100 \%$ $\approx$ -0.20) is seen (Fig. \ref{fig:7}(c)). The positive slope in this regime can be explained as due to closing of the cones with increase in $H$ (see Fig. \ref{fig:7}(d)). It is seen that $\Omega_{\text{L}}^{\text{avg}}$ reduces from about 82 deg to 65 deg in the range of 2.5 to 10.5 kOe.

In the next $H$-regime (light orange shaded), the simulated $\text{R}_{\text{sim}}^{\text{SMR long}}$ Vs. $\alpha$ signal showed a cross-over from negative to positive SMR (Fig. \ref{fig:7}(e)). Notably, the crossover from negative to positive value of simulated $(\Delta R / R_{0}) \times 100 \%$ occurs in the vicinity of 14.1 kOe (Fig. \ref{fig:7}(c)) and $\Omega_{\text{L}}^{\text{avg}}$ attains $\approx 54$ deg near the crossover (Fig. \ref{fig:7}(d)). It is to be mentioned that, in a single-domain conical phase, a pure closing of the cone-angle (i.e. an absence of tilting of the cone) results in a smooth cross-over of SMR from negative to positive values around 54 deg \cite{aqeel2016}. On the other hand, in the polycrystalline SCFO$\vert$Pt hybrid, modulation of $\text{R}_{\text{exp}}^{\text{SMR long}}$ in $\alpha$-scans shows some anomalous features in the cross-over regime. This is possibly due to presence of randomly-oriented domains. Simulated $\text{R}_{\text{sim}}^{\text{SMR long}}$ vs. $\alpha$ curves in Fig. \ref{fig:7}(e) corroborate the experimental results in the cross-over region. 

At high $H$ values (light cyan shaded region), the simulated $\text{R}_{\text{sim}}^{\text{SMR long}}$ Vs. $\alpha$ curve yields positive SMR (Fig. \ref{fig:7}(f)). In this regime dominant cone-angle closing, which persists till saturation ($H = H_{\text{sat}}$), can explain the positive SMR as well as the saturation of $(\Delta R / R_{0}) \times 100 \%$ at $H = H_{\text{sat}}$ (Fig. \ref{fig:7}(c)). The simulated $(\Delta R / R_{0}) \times 100 \%$ Vs. $H$ curve matches well with the experimental data (\textit{Measurement Set 1} and \textit{Measurement Set 2}) in all $H$-regimes except in the highest denoted by cyan color. At 100 kOe, the simulated value of $(\Delta R / R_{0}) \times 100 \%$ is $\approx 4$ times higher than the experimental value. The observed discrepancy may arise if the polycrystalline sample used in the study is not a perfectly homogenized material as presumed in the simple SMR model. 

It is worth noting that change in the helicity with reversal of the $\bm{H}$ \cite{ueda2019insights} showed no influence on the observed SMR. The experimental and computed SMR data highlight the role of tilting and closing of the cones on spin transport at SCFO$\vert$Pt interface.

\section{CONCLUSION}

We present spin transport studies on low-field, room-temperature magnetoelectric (ME) multiferroic polycrystalline \ch{Sr_3Co_2Fe_{24}O_{41}}$\vert$\ch{Pt} heterostructure. In particular, we have probed field variation of transverse conical phase at room temperature by carrying out spin transport experiments, namely longitudinal spin Seebeck effect (LSSE) and spin Hall magnetoresistance (SMR). To analyze the observed SMR data in \ch{Sr_3Co_2Fe_{24}O_{41}}, we computed SMR using a simple Hamiltonian model. Our studies elucidate the strong influence of tilting and closing of the cones on the field variation of SMR. At low magnetic fields ($H <$ 2.5 kOe), the negative SMR ($R_{\text{exp}}^{\text{SMR long}}$) and the negative slope of normalized SMR, $(\Delta R / R_{0}) \times 100 \%$ Vs. $H$ are interpreted as the result of prominent tilting of the cones in this $H$-regime. Whereas, the $H$-evolution of $(\Delta R / R_{0}) \times 100 \%$ and the sign-reversal of the SMR from negative to positive at higher fields ($H \geq$ 2.5 kOe) can be mainly attributed to the closing of the cones. We also make an important observation that the change in the helicity of the conical structure that occurs with a reversal of $H$ \cite{ueda2019insights} had no influence on the observed SMR in \ch{Sr_3Co_2Fe_{24}O_{41}}. Our LSSE Vs. $H$ data resemble dc magnetization results at 300 K. The magnetization, LSSE and SMR studies indicate that a high field is required for saturation at 300 K. Our elaborate spin transport studies demonstrate the possibility of tuning the sign and magnitude of SMR with field, in regard to the potential use of this material in magnetic insulator-based spintronic devices (e.g. SMR-based spin valves \cite{chen2013theory}). Besides, the strong ME coupling in \ch{Sr_3Co_2Fe_{24}O_{41}} at room temperature makes it a potential candidate for exploring futuristic voltage-controlled spin transport devices.

\section{ACKNOWLEDGEMENTS}

The authors acknowledge Department of Science and Technology, India for providing a project grant. Electromagnetic and spin transport measurements were carried out at the National Facility for Low Temperature and High Magnetic Field. One of the authors (PG) thank UGC India for supporting through a research fellowship.

A.W. and P.G. contributed equally to this work.

\appendix

\begin{figure*}
\includegraphics[width=1\textwidth,keepaspectratio]{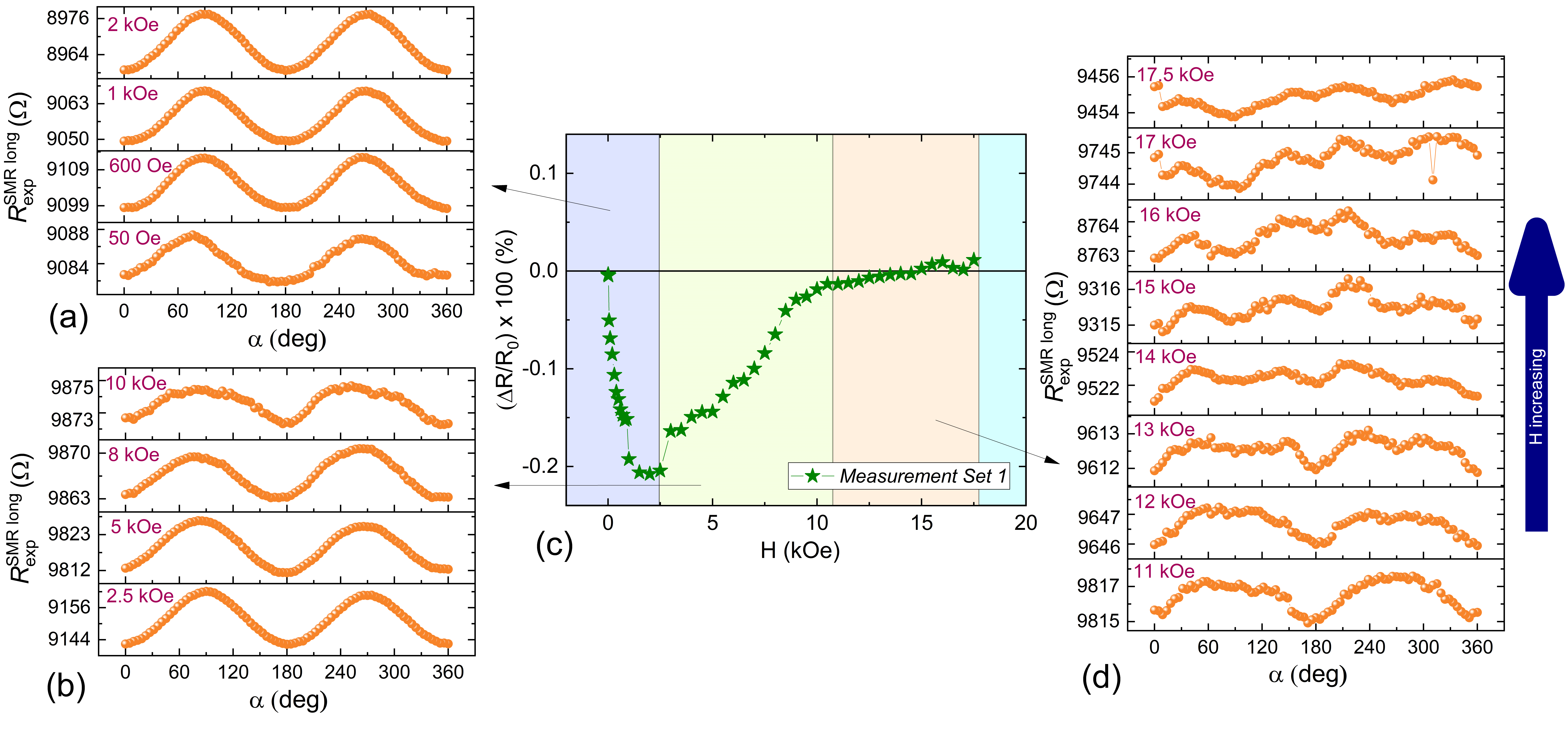}
\caption{\label{fig:S1}(a), (b), and (d) Longitudinal SMR signal ($R^{\text{SMR long}}_{\text{exp}}$) measured during $\alpha$-scan at various $H$-values in the range, 50 to 17500 Oe (\textit{Measurement Set 1} data), (c) Normalized amplitude of $R^{\text{SMR long}}_{\text{exp}}$, ($(\Delta R / R_{0}) \times 100 \% $) plotted as a function of $H$.}
\end{figure*}

\section{\label{appendix}Spin Hall magnetoresistance signal in \textit{Measurement Set 1} and \textit{2}}

\begin{figure}
\includegraphics[width=0.47\textwidth,keepaspectratio]{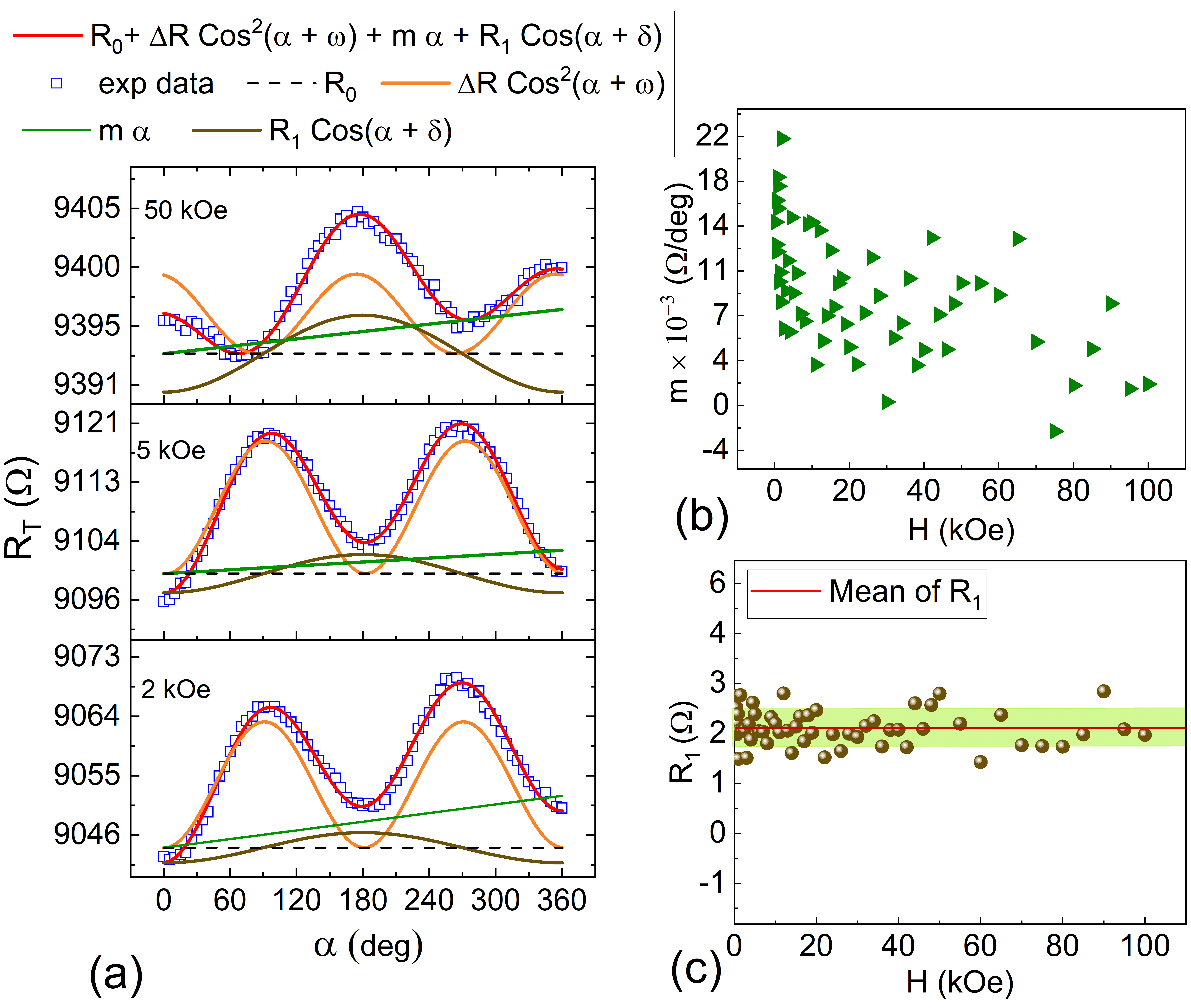}
\caption{\label{fig:S2}(a) Experimentally measured raw data (\textit{Measurement Set 2} data); $R_{\text{T}}$ Vs. $\alpha$ at three distinct $H$-values; 2, 5 and 50 kOe are fitted with various contributing terms in Eqn.\ref{EqnS1}. Fitted parameters, $m$ and $R_{1}$ representing the linear and the sinusoidal thermal drift contributions in $R_{\text{T}}$ are estimated for different $H$-values and plotted in (b) and (c), respectively.}
\end{figure}

In order to measure SMR across a wide range of magnetic field, we used two different setups; 1) a custom-built setup equipped with Lakeshore electromagnet 2T, and 2) Rotating stage holder, PPMS 14 T, Quantum Design. The custom-built set up with electromagnet is better suited for applying accurate and reliable low field values. The set up enabled us to carry out SMR studies in the field range, 50 Oe to 17.5 kOe (\textit{Measurement Set 1}). On the other hand, the rotating stage holder (PPMS 14 T) helped us explore SMR in the high field regime (\textit{Measurement Set 2}) in the range of 300 Oe to 100 kOe. Two distinct sets of Hall-bar leads were used to measure longitudinal SMR in \textit{Measurement Set 1} and \textit{Measurement Set 2}. The longitudinal resistance measured in these two sets differ in magnitude. Nevertheless, the normalized amplitude of SMR (i.e. percentage SMR) for the two sets were consistent, and in agreement with each other.

\subsection{\label{appendix_set1}\textit{Measurement Set 1}}

Our \textit{Measurement Set 1} data are shown in Fig. \ref{fig:S1}. Room temperature $R_{\text{exp}}^{\text{SMR long}}$ Vs. $\alpha$ scans measured at various fields in the range, 50 Oe to 17.5 kOe are shown in Fig. \ref{fig:S1}(a), (b) and (d). The amplitude of modulation ($\Delta R$) was extracted for each $\alpha$-scan and the percentage SMR ($(\Delta R / R_{0}) \times 100 \% $) was estimated and plotted as a function of the magnetic field (Fig. \ref{fig:S1}(c)). At the lowest applied field of 50 Oe, $R_{\text{exp}}^{\text{SMR long}}$ peaks at 90 deg and 270 deg implying that the SMR is negative in sign (Fig. \ref{fig:S1}(a)). As the field is increased ($H <$ 2.5 kOe), there is an increase in the amplitude of the negative SMR. At 2.5 kOe, $(\Delta R / R_{0}) \times 100 \% $ has a value of -0.20$\%$. At higher fields ($2.5 \text{ kOe} \leq H < 10.5 \text{ kOe}$), the SMR retains the negative sign (Fig. \ref{fig:S1}(b)). In Fig. \ref{fig:S1}(c),  the slope of $(\Delta R / R_{0}) \times 100 \%$ Vs. $H$ is negative till 2.5 kOe (denoted by light blue shaded region); thereafter, the slope becomes positive (marked by light green shaded region). With further increase in field ($10.5 \text{ kOe} \leq H < 17.5 \text{ kOe}$), a gradual cross-over from negative to positive SMR can be observed in Fig. \ref{fig:S1}(d). This field regime is described by light orange region in Fig. \ref{fig:S1}(c). The observed positive SMR at 17.5 kOe and trend in the slope of $(\Delta R / R_{0}) \times 100 \%$ Vs. $H$ indicate that the magnitude of positive SMR may increase further at higher $H$ (cyan region). This necessitates measurement at higher values of field.

\subsection{\label{appendix_set2}\textit{Measurement Set 2}}

At 300 K and higher fields, the rotating stage holder, 14 T PPMS was used (\textit{Measurement Set 2}). The results are discussed in detail in Section \ref{exp_smr} of the main text. Here, we describe how the raw data were corrected by separating the contributions originating from thermal drifts. Representative scans of three $R_{\text{T}}$ Vs. $\alpha$ curves (measured at 2, 5 and 50 kOe) and the procedure for extraction of SMR signal are shown in Fig. \ref{fig:S2}(a). Experimental data (open blue squares) at 2 and 5 kOe show negative SMR. At 50 kOe, the SMR is positive. 
Contributing factors for total measured raw $R_{\text{T}}$ Vs. $\alpha$ data are given in Eqn.\ref{EqnS1}.

\begin{equation}\label{EqnS1}
   R_{\text{T}} = R_{0} + \Delta R \cos^{2}{(\alpha + \omega)} + \text{m}\alpha + R_{1}\cos{(\alpha + \delta)} 
\end{equation}

The first and the second term represent the base resistance value of Platinum ($R_{0}$) and the SMR signal (with amplitude of modulation, $\Delta R$), respectively. It is discernible that all the $R_{\text{T}}$ Vs. $\alpha$ curves exhibit drift (increase in $R_{\text{T}}$) as the scan progresses. We perceive that it could be mainly due to a gradual increase in temperature towards the set value. Our first simple assumption is that over any $\alpha$-scan, the thermal drift is linear with angle / time. This is denoted by the third term with a linearity constant $\text{m}$. As the linear thermal-drift term was not sufficient to account for the shape of the $R_{\text{T}}$ Vs. $\alpha$ curve, an additional term of sinusoidal nature was essential. This is given by the fourth term with amplitude, $R_{1}$ and the phase, $\delta$.

Figure \ref{fig:S2}(a) shows fitting of the $R_{\text{T}}$ Vs. $\alpha$ curves with different contributions in Eqn.\ref{EqnS1}. For instance, at 2 kOe, the estimated fitting parameters are; $R_{0}$ = 9044.52 $\Omega$,  $\Delta R$ = 18.70 $\Omega$, $\omega$ = 89.0 deg, $m$ = 0.0214 $\Omega / \text{deg}$, $R_{1}$ = 2.24 $\Omega$ and $\delta$ = 0 deg. The linear temperature evolution of the resistance of Pt Hall-bar can be described by Eqn.\ref{A3}. Here, $R_{0}$ is the base resistance of Pt layer at a reference temperature and $\alpha_{\text{R}}$ is the temperature coefficient of resistance ($\alpha_{\text{R}}$ = 0.003927 $\text{K}^{\text{-}1}$ (ambient value for bulk Pt)). In order to estimate the temperature change / drift ($\Delta T$) responsible for the resistance change ($\Delta R_{\text{T}}$) in the third and the fourth terms in Eqn.\ref{EqnS1}, we use the Eqn.\ref{A3}.

\begin{equation}
  \alpha_{\text{R}}= \frac{1}{R_{0}} \text{ } \frac{\Delta R_{\text{T}}}{\Delta T}
\label{A3}    
\end{equation}

For a full $\alpha$ scan at 2 kOe, the estimated $\Delta R_{\text{T}}$ for the third and the fourth term (Refer Eqn.\ref{EqnS1}) equals to 7.7 $\Omega$  (i.e. $m \times 360$ deg) and 2.24 $\Omega$ (i.e. $R_{1}$), respectively. Replacing the $\Delta R_{\text{T}}$ value in Eqn.\ref{A3} gives $\Delta T =$ 0.22 K ($\Delta T =$ 0.06 K) for the linear thermal drift (sinusoidal thermal-drift) term. 

It is noteworthy that we carried out $R_{\text{T}}$ Vs. $\alpha$ scans from 300 Oe to 100 kOe by increasing the field to various intermittent values. As time progressed, the $\alpha$-scans recorded at increasing field-values showed reduction in the thermal drift, indicating a very gradual approach towards the stable temperature. The fitting parameters, namely $m$ and $R_{1}$ estimated for different field values are plotted as a function of field as shown in Fig. \ref{fig:S2}(b) and (c), respectively. The gradual decay of $m$ confirms the slow attainment of the set temperature. On the other hand, field-independent value of $R_{1}$ rules out any possibility of the ordinary Hall signal. Moreover, we argue that the tiny sinusoidal contribution denoted by amplitude $R_{1}$ may have originated from the design of the rotating stage holder (PPMS) as we do not observe the same in the \textit{Measurement Set 1} performed on another set up.

The experimental SMR data were corrected by removing the thermal drifts (both $m$ and $R_{1}$ contributions) from the raw data for each field. The corrected data are shown in Fig. \ref{fig:5} in the main text.

\bibliography{scfoplain}

\end{document}